\documentclass[10pt,preprint]{aastex}

\usepackage{graphicx,color}
\usepackage{natbib}
\usepackage{amssymb}

\newcommand{\eg}{{\it{e.g.}}}
\newcommand{\Atwood}{\ensuremath{\mathrm{At}}}
\newcommand\citeeg[1]{\citep[\eg{},][]{#1}}
\newcommand{\magmu}{}
\newcommand{\magH}{B}
\newcommand{\magh}{b}

\newcommand{\npku}{\ensuremath{\left ( n + k U \right )}}
\newcommand{\nmku}{\ensuremath{\left ( n - k U \right )}}
\newcommand{\dnpku}{\ensuremath{\left ( n + k U \right )'}}
\newcommand{\wonpku}{\ensuremath{\left ( \frac{w}{n + k U} \right )}}
\newcommand{\dwonpku}{\ensuremath{\left ( \frac{w}{n + k U} \right )}'}
\newcommand{\donlywonpku}{\ensuremath{\left ( \frac{w'}{n + k U} \right )}}
\newcommand{\ddwonpku}{\ensuremath{\left ( \frac{w}{n + k U} \right )}''}

\begin{document}

\title{Bubble-Wrap for Bullets: The Stability Imparted By A Thin Magnetic Layer}
\shorttitle{Stability Imparted By A Thin Magnetic Layer}
\shortauthors{Dursi}

\author{L. J. Dursi}
\affil{Canadian Institute for Theoretical Astrophysics,
                 University of Toronto, 
	 	 Toronto, ON, M5S~3H8, 
		 Canada}
\email{ljdursi@cita.utoronto.ca}

\slugcomment{In preparation, \today}

\begin{abstract}
There has been significant recent work by several authors which examines
a situation where a thin magnetic layer is `draped' over a core merging
into a larger cluster; the same process also appears to be at work at
a bubble rising from the cluster centre.  Such a thin magnetic layer
could thermally isolate the core from the cluster medium, but only if the
same shear process which generates the layer does not later disrupt it.
On the other hand, if the magnetized layer can stabilize against the
shear instabilities, then the magnetic layer can have the additional
dynamical effect of reducing the shear-driven mixing of the core's
material during the merger process.   These arguments could apply
equally well to underdense cluster bubbles, which would be even more
prone to disruption.  While it is well known that magnetic fields can
suppress instabilities, it is less clear that a thin layer can suppress
instabilities on scales significantly larger than its thickness.

We consider the here stability imparted by a thin magnetized layer.
We investigate this question in the most favourable case, that of two
dimensions, where the magnetic field can most strongly affect the stability.
We find that in this case such a layer can have a significant stabilizing
effect even on modes with wavelengths $\lambda$ much larger than the
thickness of the layer $l$ --  to stabilize modes with $\lambda \approx
10 l$ requires only that the Alfv\'en speed in the magnetized layer is
comparable to or greater than the relevant destabilizing velocity -- the
shear velocity  in the case of pure Kelvin-Helmholtz like instability,
or a typical buoyancy velocity in the case of pure Rayleigh-Taylor like
instability.  We confirm our calculations with two-dimensional numerical
experiments using the Athena code.
\end{abstract}


\section{INTRODUCTION}
\label{sec:intro}

It has been known for some decades in the space science community that
an object moving super-Alfv\'enically in a magnetized medium can very
rapidly sweep up a significant magnetic layer which is then `draped'
over the projectile \citeeg{1980Ge&Ae..19..671B}.   After recent
observations of very sharp `cold fronts' in galaxy clusters (see for
instance \cite{ShocksAndColdFrontsReview}), there has been significant
interest in applying this idea of magnetic draping in galaxy clusters
\citep[\eg{},][]{vikhlininetal2001,lyutikovdraping, asai04, asai05,
asai06} as such a magnetic field could inhibit thermal conduction
across the front \citeeg{ettorifabian} allowing it to remain sharp over
dynamically long times.

The effect of a strong draped magnetic layer could be even
greater for underdense objects, such as for bubbles moving
through the intercluster medium, as seen in many cool-core clusters
\citeeg{2005Natur.433...45M,2004ApJ...607..800B}.   In this case, the
bubble would be quickly disrupted on rising absent some sort of support
\citeeg{flashbubbles}.   However, the draping of a pre-existing magnetic
field may strongly alter the dynamics, as seen recently in simulations
\citep{ruszkowskiI, ruszkowskiII}.

However, the same shear motion which gives rise to the magnetic
draping can also drive instabilities which could then disrupt the
layer.  It is clear that magnetic fields can stabilize against shear
instabilities \citep{chandra}, and this was considered in this context
in \cite{vikhlininetal2001}.  However, in that work, the instability
was considered between two semi-infinite slabs with differing magnetic
fields -- that is, the geometric thinness of the magnetic field was
not taken into account.   This is an appropriate regime for considering
perturbations much smaller than the thickness of the layer, but less so
for modes which could disrupt the layer and contribute to the stripping
of the core.   For these modes, clearly that the layer is in fact thin
must have some effect, and naively one might expect that the thin layer
would only be effective in stabilizing modes with size comparable to
the breadth of the layer.

Here we consider the linear growth of Kelvin-Helmholtz and Rayleigh-Taylor
instabilities in the presence of a thin magnetized layer.   We consider
the instability in two dimensions, with field lines lying in the
plane and parallel to the interface, and with perturbations along the
direction of the magnetic field.   This is the case in which the layer
could most strongly influence the stability; if such a thin layer could
not stabilize the flow in such a restricted geometry, it would surely be
torn up by instabilities in the more realistic three-dimensional case.
In three dimensions, interchange modes with wavenumbers perpendicular to
the field lines are essentially unaffected by the presence of the magnetic
field, and thus the field cannot stabilize these modes.   However,
even in this case, overall mixing can be reduced by the presence of a
quite weak field \citeeg{athenamhdrt}, much weaker than expected in this
case, and the introduction of such a stark asymmetry (modes in one plane
attenuated but in a perpendicular plane unaffected) could have interesting
observable consequences.

In \S\ref{sec:derivation} we pose the problem, in \S\ref{sec:rt} and
\S\ref{sec:kh} we derive the growth rates and stability boundaries for
the Rayleigh-Taylor and Kelvin-Helmholtz instabilities, respectively,
and in \S\ref{sec:simulations} we confirm our analytic results
with two-dimensional numerical experiments using the Athena code
\citep{athenajcp}.   We conclude in \S\ref{sec:discussion}.

\section{LINEAR THEORY}
\label{sec:derivation}

\begin{figure}
\centering
\plotone{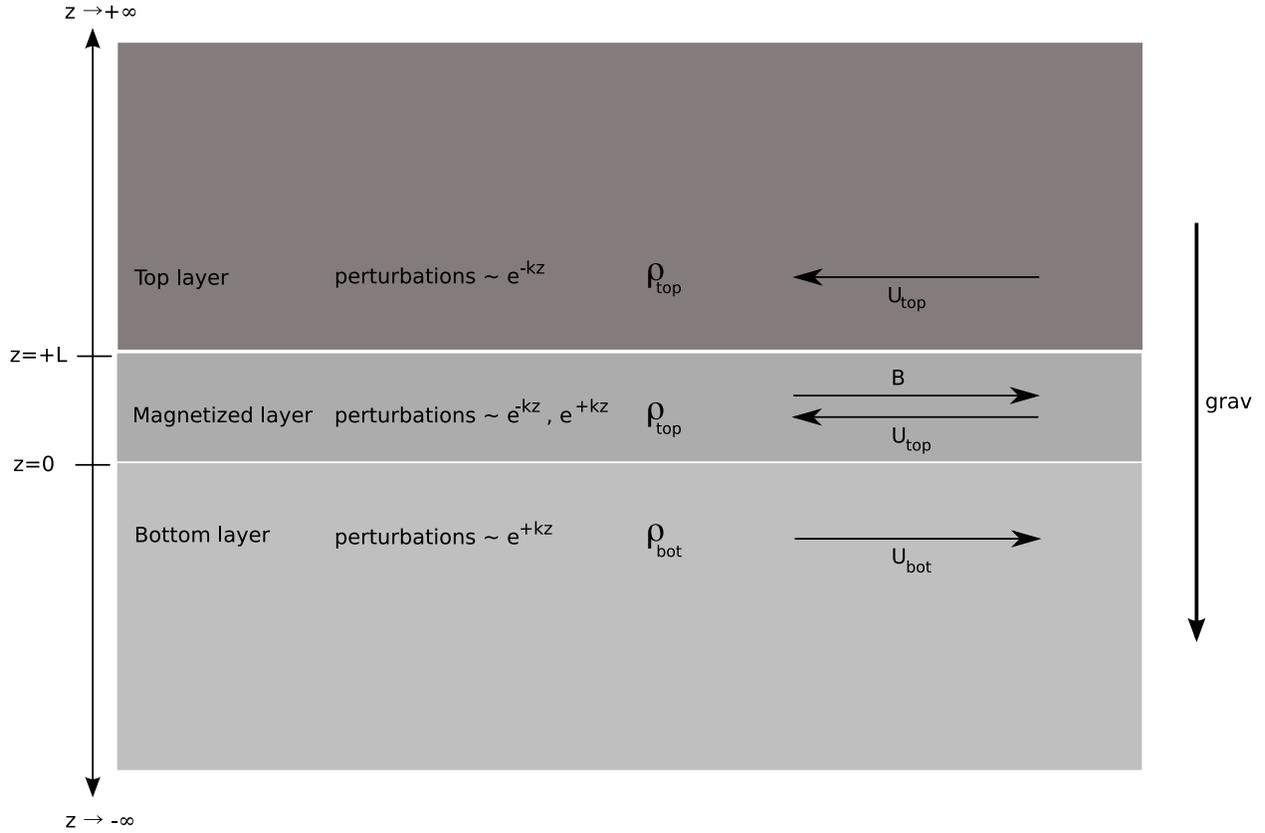}
\caption{A sketch of the problem under consideration.   We consider a
three-layer problem, with two semi-infinite, uniform, magnetic-field-free
regions separated by a layer with a horizontal magnetic field.
We consider only field (and disturbances) in the direction of the shear.
We will generally consider the magnetized layer to have the same density 
and velocity as the top fluid.   Disturbances are modeled as periodic in
the horizontal direction, and vanishing at $\pm \infty$ in the vertical direction.}
\label{fig:sketch}
\end{figure}

We follow the approach and notation of \cite{chandra}, particularly
\S~105.   We begin with the equations of two-dimensional, incompressible, inviscid,
magnetohydrodynamics:
\begin{eqnarray}
\label{eq:momentum}
\frac{\partial U_i}{\partial t} + U_j \frac{\partial U_i}{\partial x_j}
            - \frac{\magmu \magH_j}{4 \pi \rho} \left ( \frac{\partial \magH_i}{\partial x_j} - \frac{\partial \magH_j}{\partial x_i}\right) & = &
            -\frac{1}{\rho} \frac{\partial p}{\partial x_i}  + g_i \frac{\delta \rho}{\rho}\\
\label{eq:induction}
\frac{\partial \magH_i}{\partial t} + \frac{\partial}{\partial x_j} \left ( U_j \magH_i - \magH_j U_i \right ) & = & 0 \\
\label{eq:incompressible}
\frac{\partial U_i}{\partial x_i} & = & 0 \\
\label{eq:nomonopoles}
\frac{\partial \magH_i}{\partial x_i} & = & 0
\end{eqnarray}
where $x_i$ is the $i$-coordinate and can take the values $(x,z)$, $U_i$
is the velocity in the $i$-coordinate direction, $\rho$ is density,
$\delta\rho$ is any fluctuation in the density, $p$ is pressure,
$\magH$ is the magnetic field, $g$ is the gravitational acceleration
with gravity pointing `down' (\eg, in the direction of $-{\bf{\hat{z}}}$).
and summation over repeated indicies is implied.  
We consider velocity, magnetic fields, and pressure of the form
\begin{eqnarray}
\label{eq:velperts}
{\bf{U}} & = & (u + U, w) \quad u,w \ll U, \\
\label{eq:magperts}
{\bf{\magH}} & = & (\magh_x + \magH, \magh_z)
       \quad \magh_x,\magh_z \ll \magH
\end{eqnarray}
where $\magH$, $W$, and $\rho$ are constant within each region.  
We will assume that all velocities are highly non-relativistic, so that
displacement currents and relativistic effects may be neglected.  We also
assume that the flow velocities are much less than the sound
speed, and thus may consider incompressible flow.    A sketch of the
situation under consideration is shown in Fig.~\ref{fig:sketch}.

Following \cite{chandra}~\S105, we consider plane wave perturbations of the
form $e^{i ( k x + n t)}$
for $\magh_x, \magh_z, u, w, dp, \delta \rho$.   Linearizing the equations
above results in
\begin{eqnarray}
\label{eq:eq180}
i \rho \left ( n + k U \right ) u + \rho U' w - \frac{\magmu \magh_z}{4 \pi} \magH' & = & -i k dp, \\
\label{eq:eq182}
i \rho \left ( n + k U \right ) w  - \frac{\magmu \magH} {4 \pi} \left ( i k \magh_z - \magh_x' \right ) + \frac{\magmu \magh_x}{4 \pi \rho} \rho \magH' & = & dp' - g \delta \rho, \\
\label{eq:eq183x}
i \magh_x \left ( n + k U \right ) & = & i k \magH u  - w \magH' + \magh_z U',\\
\label{eq:eq183z}
i \magh_z \left ( n + k U \right ) & = & i k \magH w, \\
\label{eq:eq184}
i \delta \rho \left ( n + k U \right ) & = & -w \rho', \\
\label{eq:lincompress}
i k u + w' & = & 0, \\
\label{eq:lsolenoidal}
i k \magh_x + \magh_z' & = & 0.
\end{eqnarray}
Because we allow the magnetic field strength to vary between layers, there
are extra terms in Eqns.~\ref{eq:eq180},\ref{eq:eq182},\ref{eq:eq183x}
proportional to $\magH'$ which do not vanish.  Within the layers
themselves, these terms are identically zero; but their existence will lead to more complex
boundary conditions at the interfaces.

\subsection{Solution Within Uniform Layers}

Equations~\ref{eq:eq183x},\ref{eq:eq183z},\ref{eq:eq184}, and \ref{eq:lincompress} can be solved to
express $\magh_x, \magh_z, \magh_x', \magh_z', \delta \rho$, and $u$ in terms of $w$ and its derivatives,
the system parameters $\magH, \magH', \rho, \rho', U$, the growth rate $n$, and the wavelength of the 
disturbance $k$.  This means that we can begin to write the dispersion relation $n(k)$ in terms of 
only the perturbed z-velocity $w$ and the known parameters.   We can do this by using Eqns.~\ref{eq:eq180}
and \ref{eq:eq182} to eliminate $dp$, and then eliminate all other perturbed variables in favour of $w$, leaving
\begin{eqnarray}
2 k \frac{\magmu \magH \magH'}{4 \pi \rho} \left ( -k w U' + \left ( n + k U \right ) w'\right) + \nonumber \\
\npku^2 \left ( \frac{\rho'}{\rho} \left ( \frac{g k^2 w}{\npku} + U' k w - \npku w'\right ) + \left ( \npku \left ( k^2 w - w'' \right ) + k w U'' \right ) \right ) + \nonumber \\
- \frac{k^2}{\npku} \frac{\magmu \magH^2}{4 \pi \rho} \left ( k w \left ( \npku \left( n k + U'' \right) + k^3 U^2 - 2 k U'^2 \right ) - \left ( n + k U \right ) \left ( - 2 k U' w' + \left (n + k U\right) w'' \right ) \right ) = 0.
\label{eq:elimdp}
\end{eqnarray}

Within the three layers, where $\magH' = 0$, $U' = 0$, and $\rho' = 0$, we have
\begin{equation}
\left ( k^2 \frac{\magmu \magH^2}{4 \pi \rho} -\left ( n + k U \right)^2 \right ) \left ( k^2 w - w'' \right ) = 0
\label{eq:withinregion}
\end{equation}
which, since we are only interested in negative $n^2$, leaves us with only the solutions
\begin{equation}
w(z) \sim e^{+ k z}, \quad w(z) \sim e^{- k z}
\end{equation}
as in the case of uniform field, and as anticipated in Fig.~\ref{fig:sketch}.

\subsection{Boundary Conditions Between Layers}

The normal displacement of the interface is $- i w/(n + k U)$; this
can be seen by expressing the perturbed interface position $\delta z$
in the same form of the other perturbed variables, and the linear-order
evolution equation for $\delta z$ becomes, $i (n + k U) \delta z = w$.

Because the displacement of the interface must be unique, $w/(n + k U)$
must  be continuous across the interfaces.   This and a related quantity,
the shifted time derivative $(n + k U)$ occurs frequently enough that it
is useful to express Eq.~\ref{eq:elimdp} in terms of these quantities.
Doing so results in
\begin{eqnarray}
-2 \dnpku \dwonpku \npku \rho + \ddwonpku \left ( k^2 \frac{\magmu \magH^2}{4 \pi} - \npku^2 \rho \right)  \nonumber \\
-k^4 \frac{\magmu \magH^2}{4 \pi} \wonpku + k^2 \npku^2 \rho \wonpku + 2 k^2 \dwonpku \frac{\magmu \magH \magH'}{4 \pi} - \nonumber \\
\dwonpku \npku^2 \rho' + g k^2 \wonpku \rho' = 0
\label{eq:slhs}
\end{eqnarray}

Integrating this equation over an infinitesimal region across either of the interfaces gives us the boundary
conditions across this interface.   In doing so, terms that contain no derivatives vanish in the limit, 
and we are left with
\begin{equation}
g k^2 \left [ \rho \right ] \left . \wonpku \right |_i  - \left [ \rho \npku^2 \donlywonpku \right ]
+ k^2 \left [ \frac{\magmu \magH^2}{4 \pi} \donlywonpku \right ] = 0
\label{eq:bc}
\end{equation}
where $[f]$ indicates the jump in a quantity $f$ across the interface, and a subscript $i$ refers
to a value at the interface.
Note that when $\magH$ is constant, this reduces exactly to the homogeneous field case found in \cite{chandra}~\S106.

\subsection{Matching the Solutions}

In the top and bottom layer, one of the two solution branches ($w \sim \exp{(+k z)}$ and $w \sim \exp{(-k z)}$, respectively)
are clearly unphysical.   In the middle layer, however, both can coexist, and so we have as forms for the solutions
\begin{equation}
w = \left \{ \begin{array}{ll}
            w_3 e^{-k z} & \mathrm{top\,layer} \\
            w_{2+} e^{+k z} + w_{2-} e^{-k z} & \mathrm{middle\, layer} \\
            w_1 e^{+k z} & \mathrm{bottom\, layer}
	    \end{array} \right .
\end{equation}

Because $w/(n + k U)$ must be continuous across the interface, we have
\begin{eqnarray}
\frac{w_1}{\npku} & = & \frac{w_{2+} + w_{2-}}{\nmku}, \\
\frac{w_3 e^{-k l}}{\nmku} & = & \frac{w_{2+} e^{+k l} + w_{2-} e^{-k l}}{\nmku}.
\end{eqnarray}
These can be solved for the components of the intermediate velocity in terms of the outer layer
velocities, giving
\begin{eqnarray}
w_{2+} & = & \frac{w_1 \nmku - w_3 \npku}{(e^{2 k l} - 1) (n + k U)}, \\
w_{2-} & = & \frac{w_1 e^{2 k l} (n - k U) - w_3 (n + k U)}{(e^{2 k l} - 1) (n + k U)}. \\
\dwonpku_{2, z=l} & = & w_1 k \frac{e^{k l}}{e^{2 k l}-1} \frac{-n + k U}{n - k U} + w_3 k \frac{e^{-k l} \left( e^{2 k l} + 1\right)}{e^{2 k l} - 1}\\
\dwonpku_{2, z=0} & = & 2 w_1 k \frac{e^{2 k l}+1}{e^{2 k l}-1} \frac{-n + k U}{n + k U} + 2 w_3 k \frac{1}{e^{2 k l}-1} 
\end{eqnarray}

We now have two boundary conditions to satisfy --- Eq.~\ref{eq:bc} at the two interfaces between the layers.

The top interface condition gives us
\begin{equation}
 \rho_{\mathrm{top}} \left ( w_3 k + \nmku \dwonpku_{2,z=l} \right) - k^2 \frac{B^2}{4 \pi} \dwonpku_{2,z=l} = 0
\end{equation}
and the bottom interface gives us
\begin{equation}
g k^2 \left ( \rho_{\mathrm{top}} - \rho_{\mathrm{bot}} \right ) \frac{w_1}{\npku} 
- \left( \rho_{\mathrm{top}} \nmku \dwonpku_{2,z=0} - \rho_{\mathrm{bot}} w_1 k \right )
+ k^2 \frac{B^2}{4 \pi} \dwonpku_{2,z=0} = 0
\end{equation}

This gives us two equations in terms of $w_1$ and $w_3$.    Using our expressions for $w_{2\pm}$ and  then solving the first equation for $w_3$ in terms of $w_1$
and substituting the result into the second gives us our final dispersion relation of
\begin{eqnarray}
\left[ n^2 + k^2 U^2 + \Atwood k \left ( g - 2 n U \right ) \right ] & + & \nonumber \\
\left (e^{2 k l} -1 \right ) \frac{\left(\nmku^2 - \frac{1}{2} k^2 v_A^2 \right) \left ( n^2 + k^2 U^2 + \Atwood k \left ( g - 2 n U \right) - \frac{1+\Atwood}{2} k^2 v_A^2 \right)}{\nmku^2 - k^2 v_A^2} & = & 0
\end{eqnarray}
where the Atwood number, $\Atwood$, is defined to be the non-dimensional density difference
\begin{equation}
\Atwood = \frac{\rho_\mathrm{top}-\rho_\mathrm{bot}}{\rho_\mathrm{top}+\rho_\mathrm{bot}}.
\end{equation}

It is worth noting that in the no magnetization limit $v_A \rightarrow 0$, the only term containing $l$ can be divided out, so that the solution
does not depend on $l$; this is as it must be, as without a magnetic field nothing distinguishes the $l$-thick middle layer from the top layer.
Also, in the infinitely thin layer limit $\exp{2 k l} \rightarrow 1$ and the solution reduces to the non-magnetized case.

\section{EFFECT ON THE RAYLEIGH-TAYLOR INSTABILITY}
\label{sec:rt}

\begin{figure}
\centering
\plotone{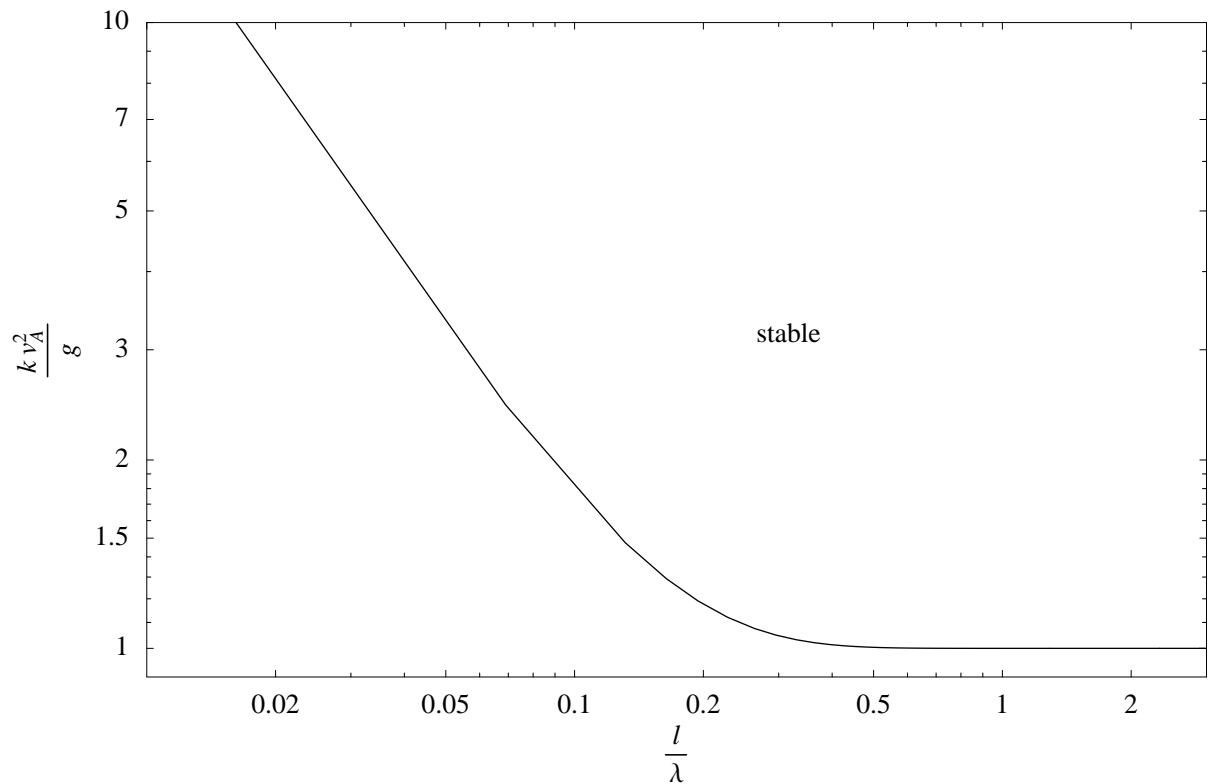}
\caption{
Plotted is the Rayleigh-Taylor stability boundary for the most-unstable
($\Atwood = 1$) case, as given in Eq.~\ref{eq:rtstable}.   The magnetic
field strength in the magnetized layer (here expressed as the Alfv\'en
speed squared in units of $g/k$) necessary to stabilize against a mode
is plotted as a function of the thickness of the layer in units of
the wavelength of the mode.    Thus, to stabilize modes an order of
magnitude greater longer than the layer is thick ($l/\lambda \approx
0.1$), a magnetic field such that $v_A^2 \gtrsim 2 g/k$ is necessary.
}
\label{fig:rtstable}
\end{figure}

If we consider the `pure' Rayleigh-Taylor instability, with no horizontal shear, then $U \rightarrow 0$ and we are left with
\begin{equation}
n^2 + \Atwood g k + \left (e^{2 k l} -1 \right ) \frac{\left(n^2 - \frac{1}{2} k^2 v_A^2 \right) \left ( n^2 + \Atwood g k - \frac{1+\Atwood}{2} k^2 v_A^2 \right)}{n^2 - k^2 v_A^2} =  0.
\label{eq:rtdispersion}
\end{equation}
This equation relates the growth rate $n$ to two other inverse
timescales on the scale of $k^{-1}$ -- an Alfv\'en frequency $\omega_A^2 =
k^2 v_A^2$, and a gravitational timescale $\tau_g^{-2} = \Atwood g k$.
Expressing the growth rate and Alfv\'en frequency in units of the inverse
gravitational timescale, and ignoring the trivial Alfv\'en wave solution 
$n = \omega_A$, the result is a quadratic in $n^2$:
\begin{equation}
e^{2 k l} \tilde n^4 + \left ( \left(1 - \omega_A^2 \right) + \left( e^{2 k l}-1 \right) \left ( 1 - \frac{2 + \Atwood}{2} \omega_A^2 \right) \right) \tilde n^2
 - \omega_A^2 \left( 1 + \frac{1}{2} \left ( e^{2 k l} - 1 \right) \left (1 - \frac{1+\Atwood}{2} \omega_A^2 \right)\right) = 0.
\end{equation}
To consider the degree to which the magnetized layer stabilizes against
the Rayleigh-Taylor instability, we consider (for simplicity) the maximally unstable case where $\Atwood \rightarrow 1$.
For stability, it is necessary and sufficient that the roots of the quadratic in $n^2$ be positive and real; 
a quadratic of the form $a x^2 + b x + c = 0$ has positive and real roots for $c/a > 0$, $b/a < 0$, and $b^2-4ac > 0$.
The first two of these conditions reduce to
\begin{eqnarray}
\omega_A^2 & > & \frac{e^{2 k l} + 1}{e^{2 k l} - 1} \\
\omega_A^2 & > & \frac{2 e^{2 k l}}{3 e^{2 k l} - 1},
\end{eqnarray}
and the third is satisfied for all real $\omega_A^2$.     Of these conditions, the first controls, as the second simply posts a
lower limit for $\omega_A^2$ of between $2/3$ and $1$, while the lower limit for the first is always greater than $1$.   Thus
the condition for stability in the case of a Rayleigh-Taylor instability with $\Atwood = 1$, which would otherwise always
be unstable, is
\begin{equation}
v_A^2 > \left ( \frac{e^{2 k l} + 1}{e^{2 k l} - 1} \right )  g k^{-1}
\label{eq:rtstable}
\end{equation}
with no strength of magnetic field able to stabilize in the case of an infinitely thin layer.   This stability criterion is shown in Fig~\ref{fig:rtstable}.

\section{EFFECT ON THE KELVIN-HELMHOLTZ INSTABILITY}
\label{sec:kh}

\begin{figure}
\centering
\plotone{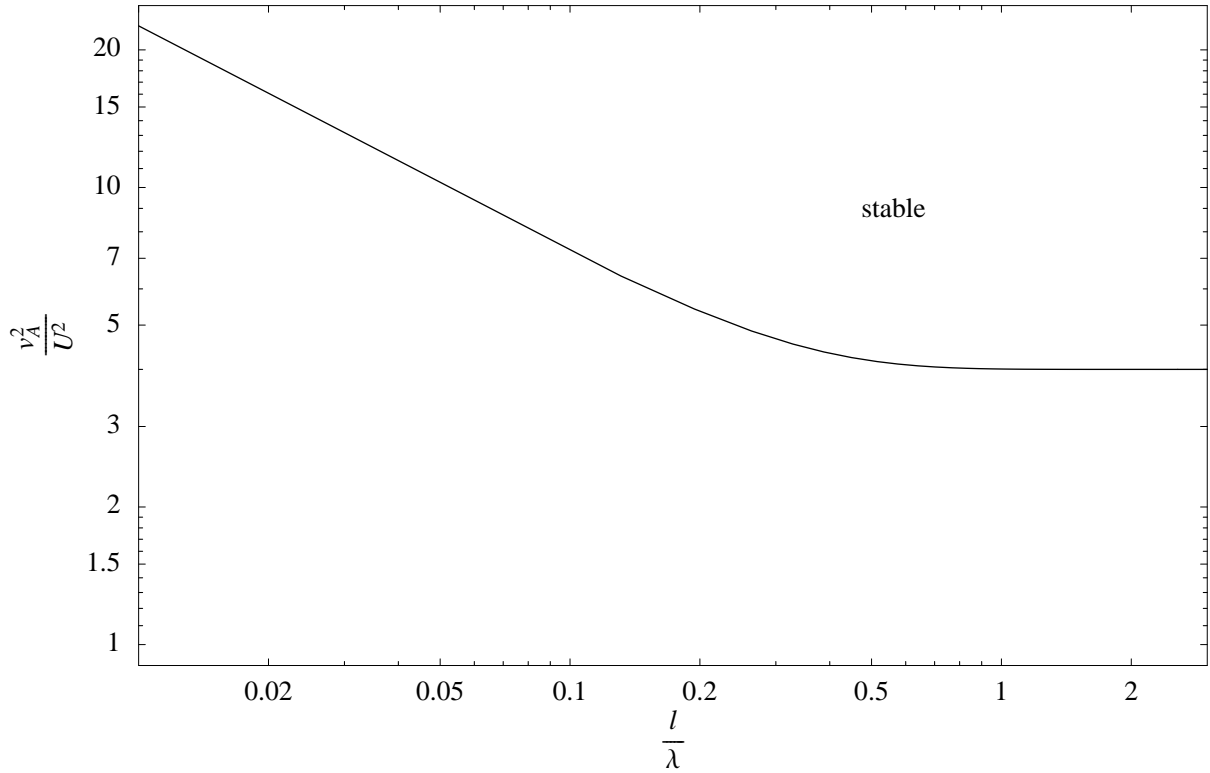}
\caption{
Plotted is the Kelvin-Helmholtz stability boundary for the most-unstable
($\Atwood = 1$) case, as given in Eq.~\ref{eq:khstable}.   The magnetic field strength in the magnetized
layer (here expressed as the Alfv\'en speed squared in units of the half-shear velocity squared, $U^2$)
necessary to stabilize against a mode is plotted as a function of the
thickness of the layer in units of the wavelength of the mode.    Thus,
to stabilize modes an order of magnitude greater longer than the layer
is thick ($l/\lambda \approx 0.1$), a magnetic field such that $v_A^2
\gtrsim 2 (2 U)^2$ is necessary.}
\label{fig:khstability}
\end{figure}

The shear terms in the case of the Kelvin-Helmholtz instability make the dispersion relation significantly
more complicated,
\begin{eqnarray}
 n^4 - 2 \left(1 + \Atwood \right ) k U  n^3 + \frac{1}{2} e^{-2 k l} \left ( \Atwood k^2 v_A^2 + e^{2 k l} \left(4 \left(1 + 2 \Atwood\right) k^2 U^2 - (2 + \Atwood) k^2 v_A^2 \right)\right)  n^2 + \nonumber && \\
- e^{-2 k l} k U \left( k^2 v_A^2 + e^{2 k l} \left( 2 ( 1 + \Atwood) k^2 U^2 - (1 + 2 \Atwood) k^2 V_A^2 \right ) \right ) n + \nonumber && \\
\frac{1}{4} e^{-2 k l} \left( 2 \Atwood k^2 U^2 k^2 v_A^2 - \left( 1 + \Atwood \right) k^4 v_A^4 + e^{2 k l} \left( 2 k^2 U^2 - k^2 v_A^2\right)\left( 2 k^2 U^2 - \left( 1 + \Atwood \right) k^2 v_A^2 \right ) \right ) & = & 0.
\label{eq:khdispersion}
\end{eqnarray}
For considering the stability boundary we will again consider the most unstable case, in this case $\Atwood = 0$.  As in the previous case, the growth rate and the Alfv\'en frequency can be expressed in terms of the other timescale of the problem, the advection time across the wavelength of the perturbation, $\tau_U = 1/(k U)$, leaving us with
\begin{eqnarray}
\tilde n^4 - 2 \tilde n^3 - \left(\omega_A^2 - 2\right) \tilde n^2 + \left(\omega_A^2 \left(1 - e^{-2 k l}\right) - 2\right) \tilde n  && \nonumber \\
+ \frac{1}{4} \left ( \left(\omega_A^2 - 2 \right)^2 - \omega_A^4 e^{-2 k l} \right) & = & 0.
\end{eqnarray}
While this is a fairly unpleasant quartic in $n$, it is a fairly approachable quadratic in $\omega_A^2$:
\begin{equation}
\frac{1}{4} \left(e^{2 k l} - 1\right) \omega_A^4  + \left (1 + \tilde n \left (e^{-2 k l} - 1 \right)  + \tilde n^2 \right) \omega_A^2 - (\tilde n - 1)^2 (\tilde n^2+1) = 0.
\end{equation}
The two roots are
\begin{equation}
\omega_A^2 = 2 \frac{ \tilde n + e^{2 k l} + (\tilde n-1)n e^{2 k l} \pm \sqrt{\left(\tilde n^2 + e^{2 k l}\right) \left(1 + \tilde n^2 e^{2 k l} \right) }}{e^{2 k l}-1} .
\end{equation}
For stability, we consider the neighborhood around $Im(\tilde n) = 0$.  In this
case, both these branches have minima for purely oscillatory modes around
$\tilde n = 1$; but in this neighborhood the second, negative, branch, has no
real solutions for $\omega_A$ with $Im(\tilde n) \neq 0$, so cannot be relevant to
the question of stability.  The positive branch has a minimum of
\begin{equation}
\omega_A^2 = 4 \frac{e^{2 k l} + 1}{e^{2 k l} - 1}
\end{equation}
so that stability is ensured when this condition is met, or
\begin{equation}
v_A^2 \ge \left ( \frac{e^{2 k l} + 1}{e^{2 k l} - 1} \right ) (2 U)^2,
\label{eq:khstable}
\end{equation}
which is the same condition for stability of the Rayleigh-Taylor instability,
but with $2 U$ replacing $g/k$; recall, however, that the two conditions 
are for two different values of the Atwood number, such that the instability
is maximized in each case.   This stability criterion is plotted in Fig.~\ref{fig:khstability}.

\section{NUMERICAL RESULTS}
\label{sec:simulations}

To confirm the results of the previous sections, numerical experiments
were performed in two dimensions using version 3.0 of the Athena code
\citep{athenajcp}, a dimensionally unsplit, highly configurable MHD code.
For the results in this work we used the ideal gas MHD solver with an
adiabatic equation of state ($\gamma = 1.4$), and the 3rd-order
accurate solver using a Roe-type flux function.   We considered a domain of
size, in code units, of $[-1/6,1/6] \times [-1/6,1/6]$ with resolution
$400 \times 400$ for the Kelvin-Helmholtz instability simulations,
and with a slightly vertically extended domain ($[-1/6,1/6] \times
[-1/4,1/4]$, $400 \times 600$) for the Rayleigh-Taylor instability
experiments.

To ensure that the resolution used was adequate, a resolution study
was performed on a fiducial run (a Rayleigh-Taylor simulation with
$B_{x,0} = 0.07, v_A = 0.0495$; the magnetic field code units are
such that the Alfv\'en speed, $v_A^2 = B^2/2$) with resolution
varying between a factor of two less than this resolution and a
factor of two more; measured growth rates varied by only approximately
$\pm 3$~percent.

The analytic results presented in previous sections were in the
incompressible limit; in our numerical experiments here the fiducial
density was $1$ in code units and the pressure was set so that the sound
speed $c_s$ would be $1$ in code units, which is an order of magnitude
larger than the velocities achieved in either set of simulations;  thus
the Mach number ${\cal{M}} < 0.1$, and incompressibility remains a reasonable
approximation.  In both
set of simulations, a magnetized layer of thickness $1/60$ was initialized
starting at $y=0$, with strength of magnetic field varied from run to run.
To keep the initial conditions in pressure equilibrium, the thermal
pressure was reduced in this layer, but because of the large sound speed
(and consequently large plasma $\beta$) this was a small reduction
(never more than a few percent).

\begin{figure}
\centering
\plotone{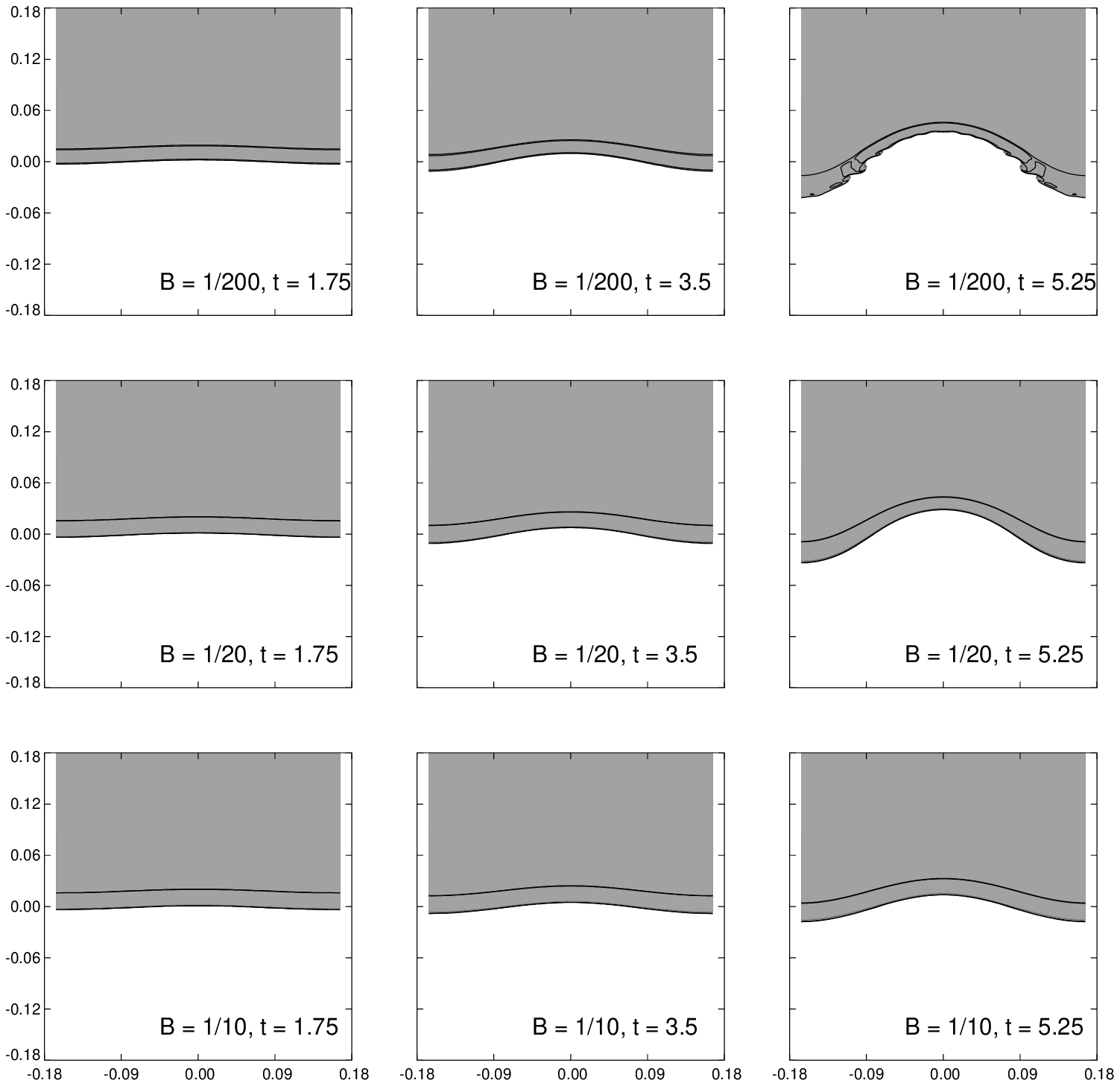}
\caption{Snapshots from representative runs of the Rayleigh-Taylor
instability with a magnetized layer, performed with the Athena code.
Shading represents density, and contours indicate magnetic field strength.
In code units, the density of the top fluid is $\rho_t = 2$, and that
of the bottom is $\rho_t = 1$ ($\Atwood = 1/3$), and pressure is set
such that in the bottom layer the adiabatic sound speed $c_s$ is 10,
and the domain is in total pressure equilibrium.  The wavelength of
the initial perturbation is $1/3$, and the thickness of the magnetized
layer is $1/60$; the acceleration due to gravity $g$ is $1/10$ and
in the negative $y$ direction.  Shown are runs with varying initial
horizontal magnetic fields, with, from top to bottom, $B_{x,0} = 1/200,
1/20, 1/10$.  Snapshots are shown at times in code units of, left to
right, approximately $(1.75, 3.5, 5.25)$.  The contours are for magnetic
field strength of $B_{x,0}/4$ and $3 B_{x,0}/4$.}
\label{fig:rtframes}
\end{figure}

\begin{figure}
\centering
\plotone{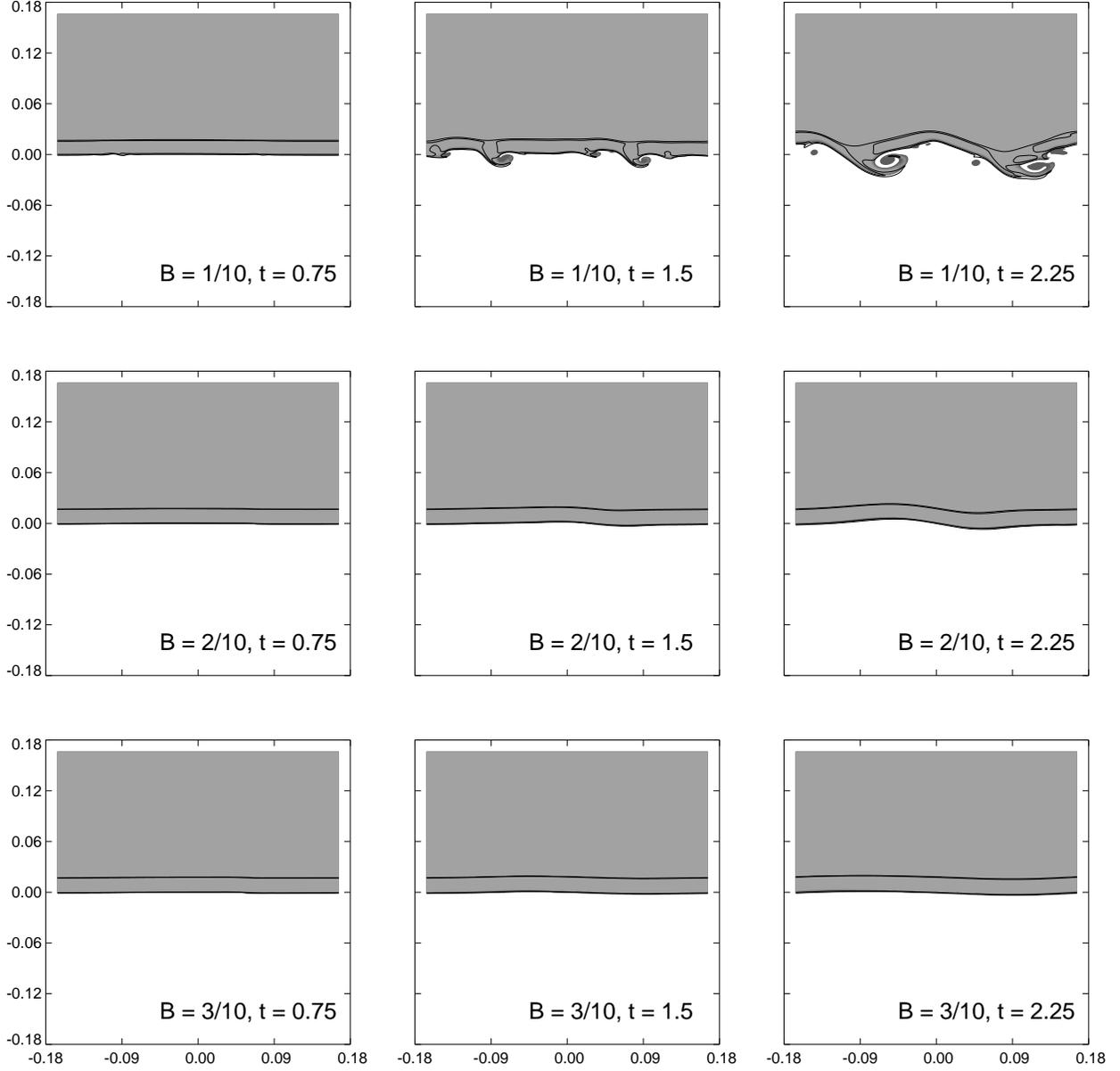}
\caption{Snapshots from representative runs of the Kelvin-Helmholtz
instability with a magnetized layer, performed with the Athena code.
Shading represents density, and contours indicate magnetic field strength.
In code units, the density of the top fluid is $\rho_t = 2$, and that
of the bottom is $\rho_t = 1$ ($\Atwood = 1/3$), and pressure is set
such that in the bottom layer the adiabatic sound speed $c_s$ is 1,
and the domain is in total pressure equilibrium.  The wavelength of
the initial perturbation is $1/3$, and the thickness of the magnetized
layer is $1/60$; the acceleration due to gravity $g$ is $1/10$ and
in the negative $y$ direction.  Shown are runs with varying initial
horizontal magnetic fields, with, from top to bottom, $B_{x,0} = 0.1,
0.2, 0.3$.  Snapshots are shown at times in code units of, left to right,
approximately $(0.75, 1.5, 2.25)$.  The contours are for magnetic field
strength of $B_{x,0}/4$ and $3 B_{x,0}/4$.}
\label{fig:khframes}
\end{figure}

In both sets of simulations, the interface was given a sinusoidal velocity
with a wavelength equal to the size of the box, and an amplitude of
$v_{pert} \approx 0.0025$.    In the Kelvin-Helmholtz case, a background
shear velocity of $-0.1$ in the $x$ direction was applied to the top
layer and the magnetized layer, and of $+0.1$ in the bottom layer.
In the Rayleigh-Taylor case, a gravitational acceleration of $g = 0.1$
was applied in the negative $y$ direction.   Both set of simulations
used an Atwood number $\Atwood = 1/3$, so that the top layer had density
$\rho_t = 2$.    Snapshots of the evolution of the simulations
are shown in Figs.\ref{fig:rtframes} and \ref{fig:khframes}.

\begin{figure}
\centering
\plotone{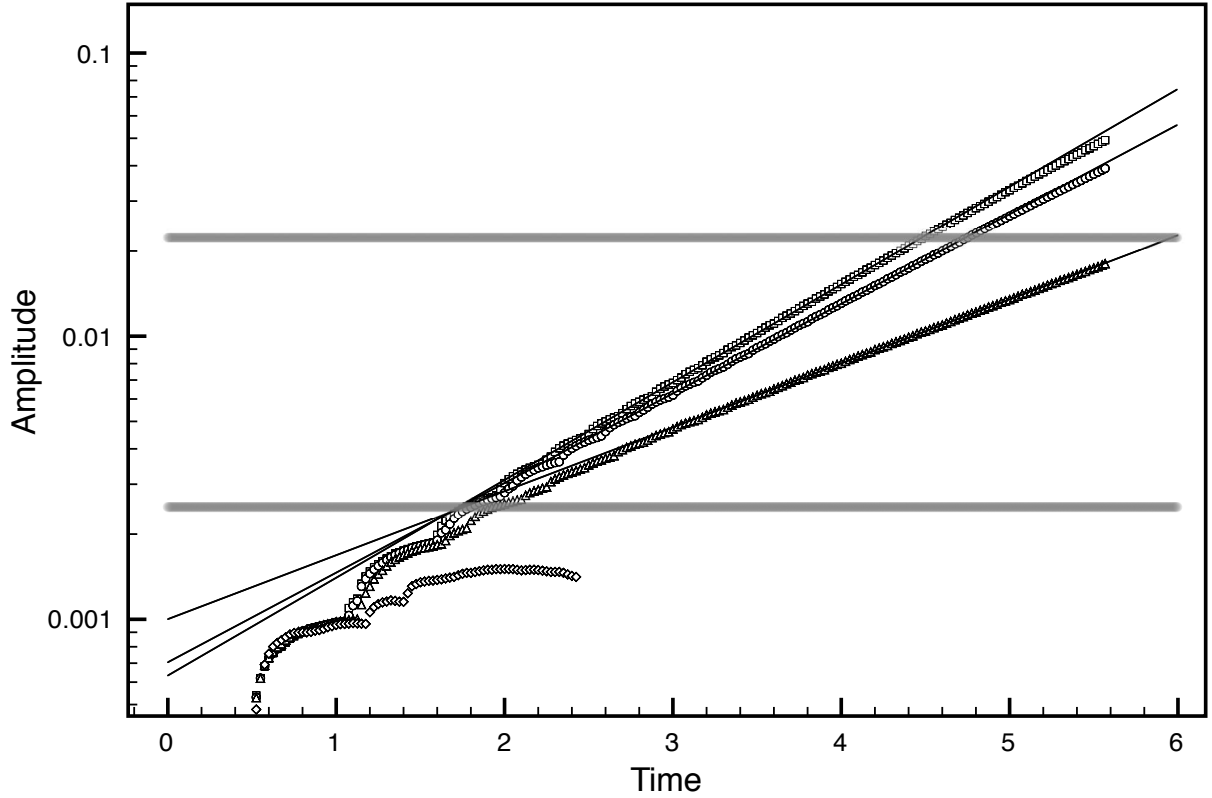}
\caption{Evolution of amplitude of the Rayleigh-Taylor perturbation with
time, for several runs (plotted are, top to bottom, $B_0 = (0.005, 0.05,
0.1, 0.2)$), shown as symbols; also plotted here are best-fit exponentials
to measure the growth rate of the instability.   Exponentials were fit
to data from those times where the amplitude was resolved by at least
three zones and where the amplitude was less than $1/15$ of the wavelength
(\eg{}, before nonlinear evolution begins to matter); for the simulations
reported here, this means the fit was performed with amplitudes in the
range $[1/400, 1/45]$, covering approximately a decade in amplitude.
The range considered is indicated by thick shaded lines; below the bottom
line the `jumpiness' due to poor resolution on the grid is visible,
and above deviations from exponential behaviour due to nonlinear effects
becomes evident.}
\label{fig:rtamplitudes}
\end{figure}

\begin{figure}
\centering
\plotone{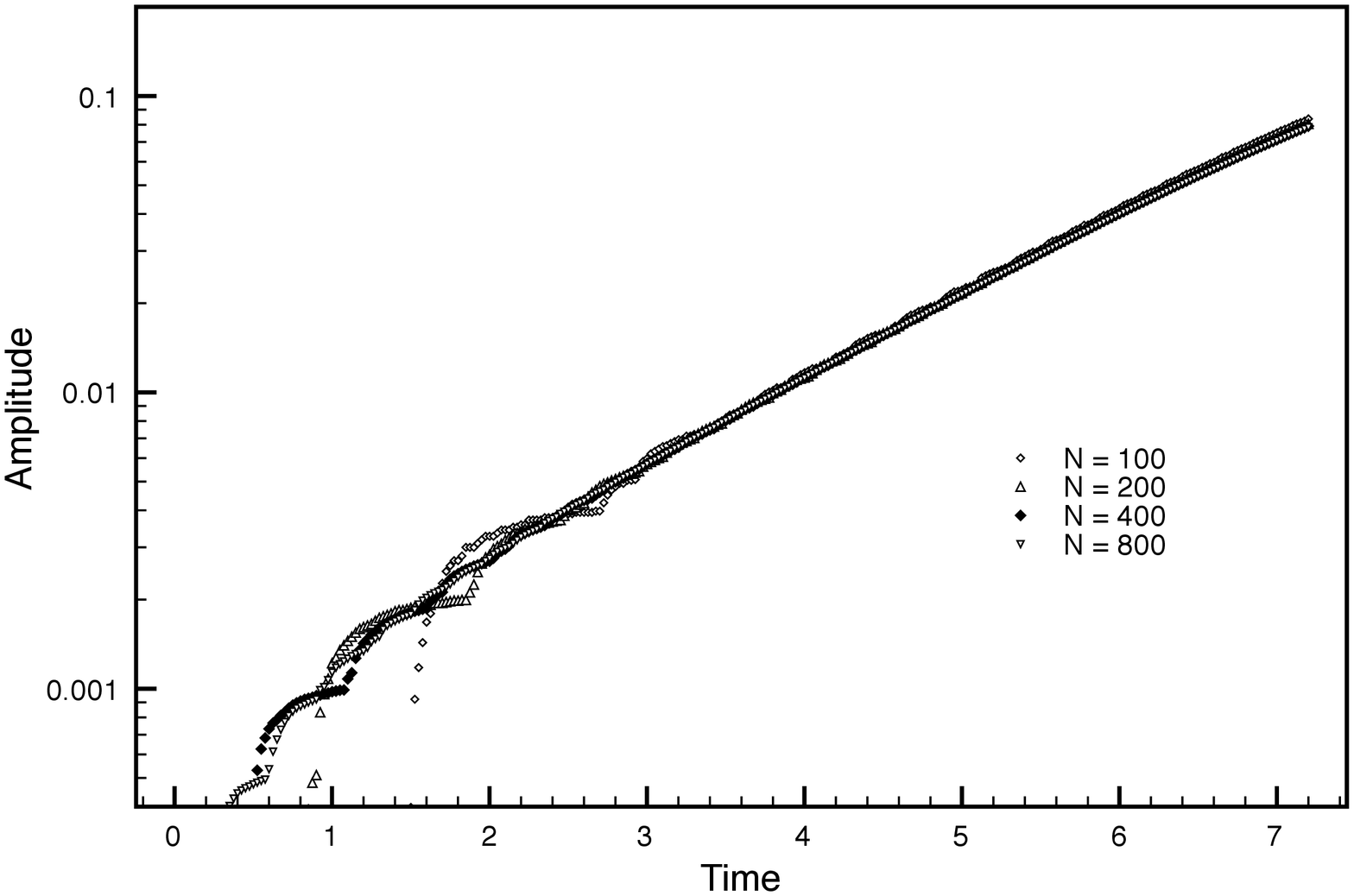}
\caption{As in Figure~\ref{fig:rtamplitudes}, but with $B_0 = 0.07$, and
varying the resolution.   Shown are for several runs with $N$ varying by a
factor of $4$, where $N$ is the horizontal resolution in the simulation,
with vertical resolution kept proportionate; thus N is the number of
points used to represent one wavelength, with the thickness of the layer
being represented by $N/20$ points.   $N = 100$ is not quite enough
resolution to accurately measure the growth rate, but the difference 
in fitted exponential growth rates between $200$, $400$, $600$, and $800$ is only a few percent.}
\label{fig:rtresstudyamplitudes}
\end{figure}

From the outputs of the simulations, growth rates for the instabilities
were calculated by considering the growth of the amplitude (measured by
finding the mean vertical position of the magnetized layer and fitting
to a sinusoid of the wavelength of the perturbed mode) as a function of time.
Exponentials were fit to this series of amplitudes,  for those times
where the amplitude was resolved by at least three zones and where the
amplitude was less than $1/15$ of the wavelength (\eg{}, before nonlinear
evolution begins to matter).  For the simulations reported here, this
means the fit was performed with amplitudes in the range $[1/400,
1/45]$, covering approximately a decade in amplitude.   Plotted in
Fig.~\ref{fig:rtamplitudes} are four examples of this procedure for the
Rayleigh-Taylor simulations.   For comparison, the results of a resolution 
study of a fiducial Rayleigh-Taylor case is shown in Fig.~\ref{fig:rtresstudyamplitudes}.

\begin{figure}
\centering
\plotone{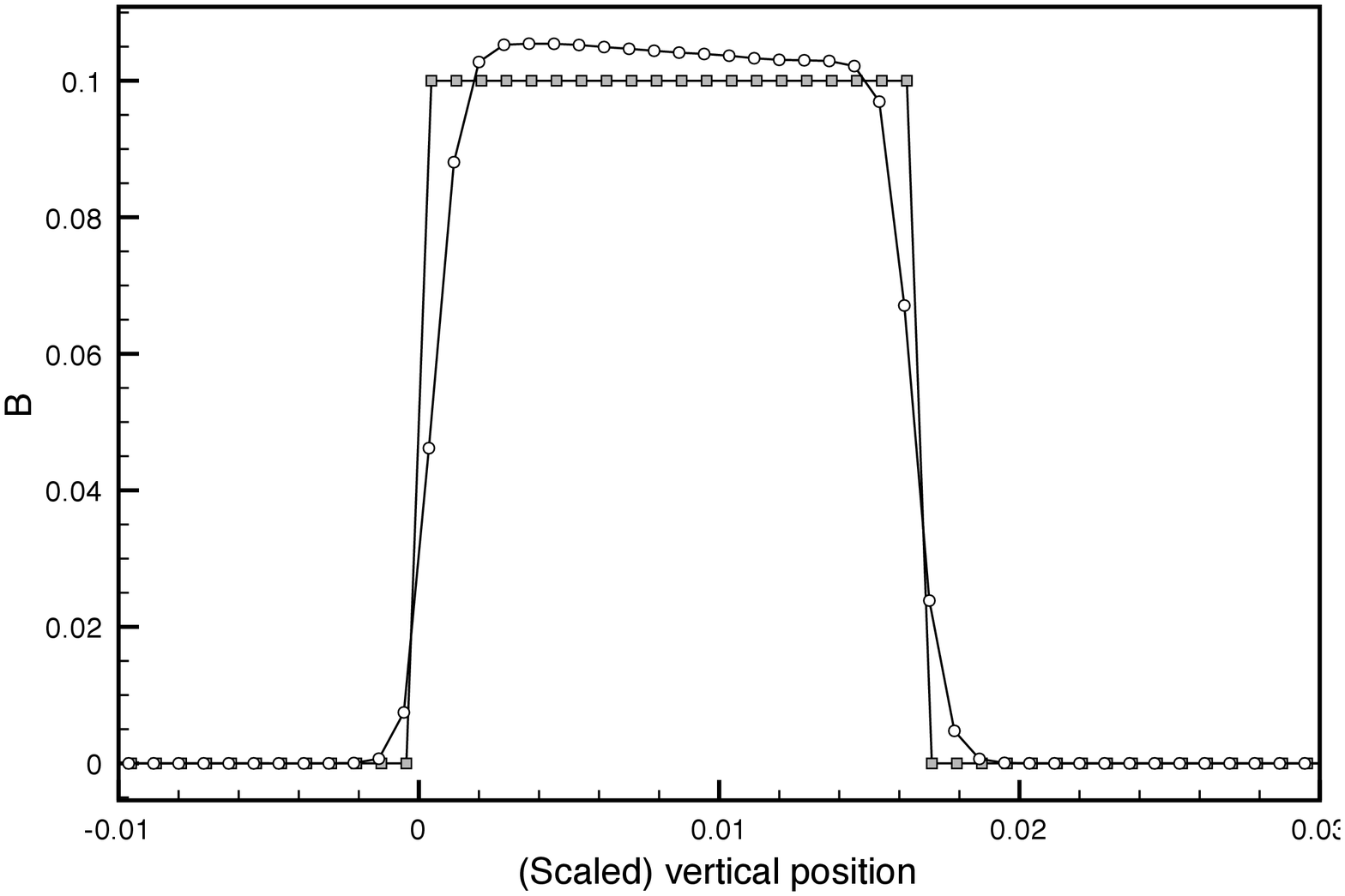}
\caption{Magnetic field profile through $x=0$ for the $B_{x,0}=1/10$
Rayleigh-Taylor simulation at time 0 (filled squares) and 3.75 (open circles).  The
later-time profile is shifted leftwards by $\delta y = 0.00675$ for
comparison.  
During the course of the simulation, the thickness of the magnetic field layer
at its peak is reduced to about 18 grid points from the original 20.   Besides
the inevitable effects of numerical diffusion slightly smearing out the profile
over approximately 18000 timesteps, effects
from the slight compressibility of the flow, and asymmetry from the direction of
gravity (to the left in this plot) is shown by the slight increase in magnetic
field density and the slope at late times.}
\label{fig:bprofile}
\end{figure}

Because of the low speeds of these flows, many timesteps (typically
on order 20000) must be take to evolve these instabilities.  Over
that period of time modest amounts of numerical diffusion slightly
modify the structure of the magnetic layer, as does the still
slightly compressible flow itself; the change is shown in
Fig.\ref{fig:bprofile}.  The growth rate of the instabilities is
very sensitive to the thickness of the layer, and this modification
of the magnetic layers thickness must be taken into account when
comparing with analytic results.

\begin{figure}
\centering
\plotone{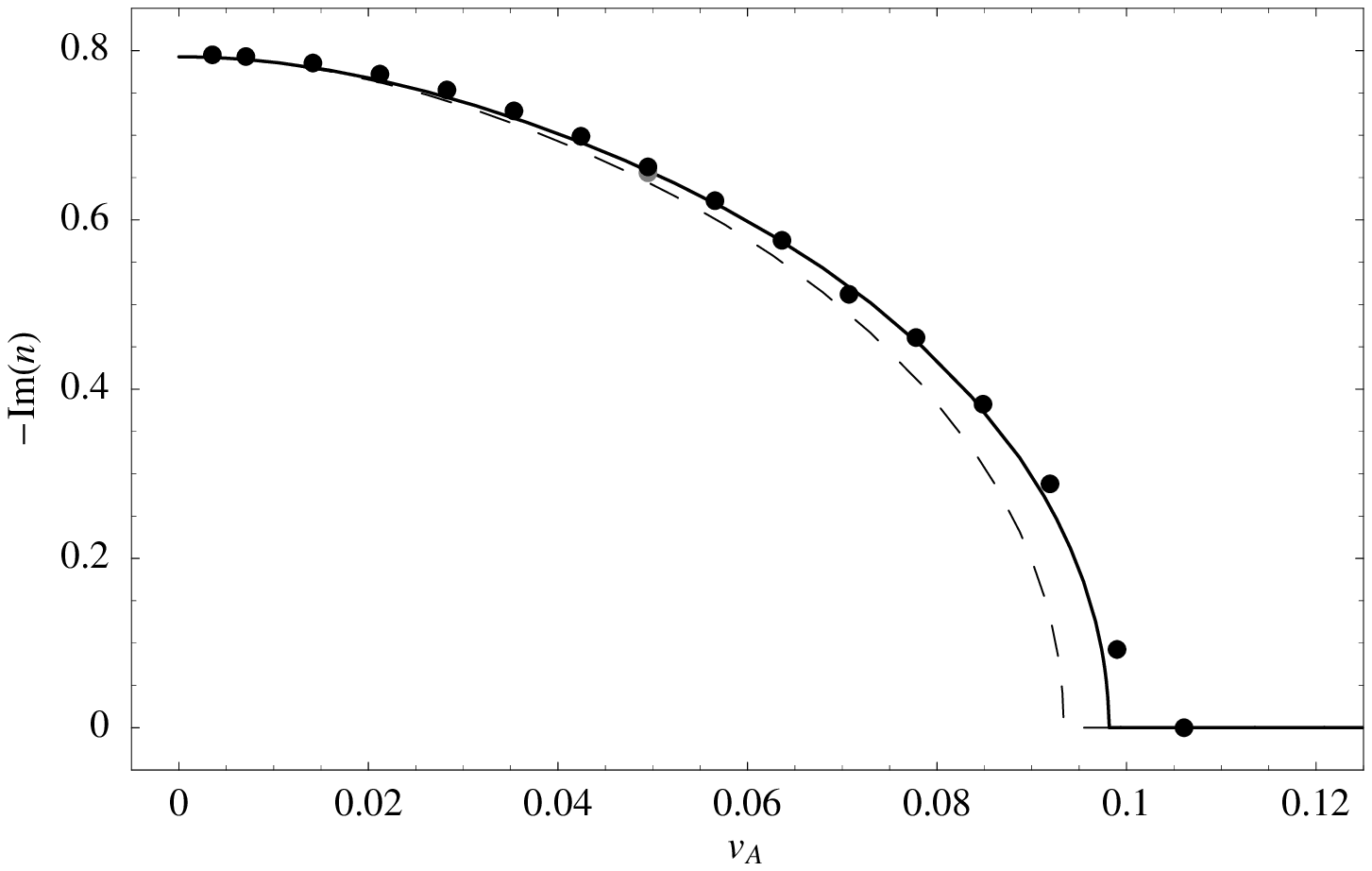}
\caption{
Rayleigh-Taylor growth rates as a function of magnetic field strength
(given here in terms of Alfv\'en speed, $v_A$) in the magnetized
layer, measured from simulation (diamonds) and predicted from
analytic theory (lines).  The dashed line gives the prediction using
the nominal width of the magnetized field layer, and the solid line
gives the prediction for the actual width of the magnetized layer
in the simulations after diffusion eats away at the profile as shown
in Fig.~\ref{fig:bprofile}.   A gray point indicates a run done at twice
the resolution to check to see if the default resolution was adequate.
The theoretical curves are taken from numerical solution of the dispersion relation, Eq.~\ref{eq:rtdispersion}.
}
\label{fig:rtgrowth}
\end{figure}

\begin{figure}
\centering
\plotone{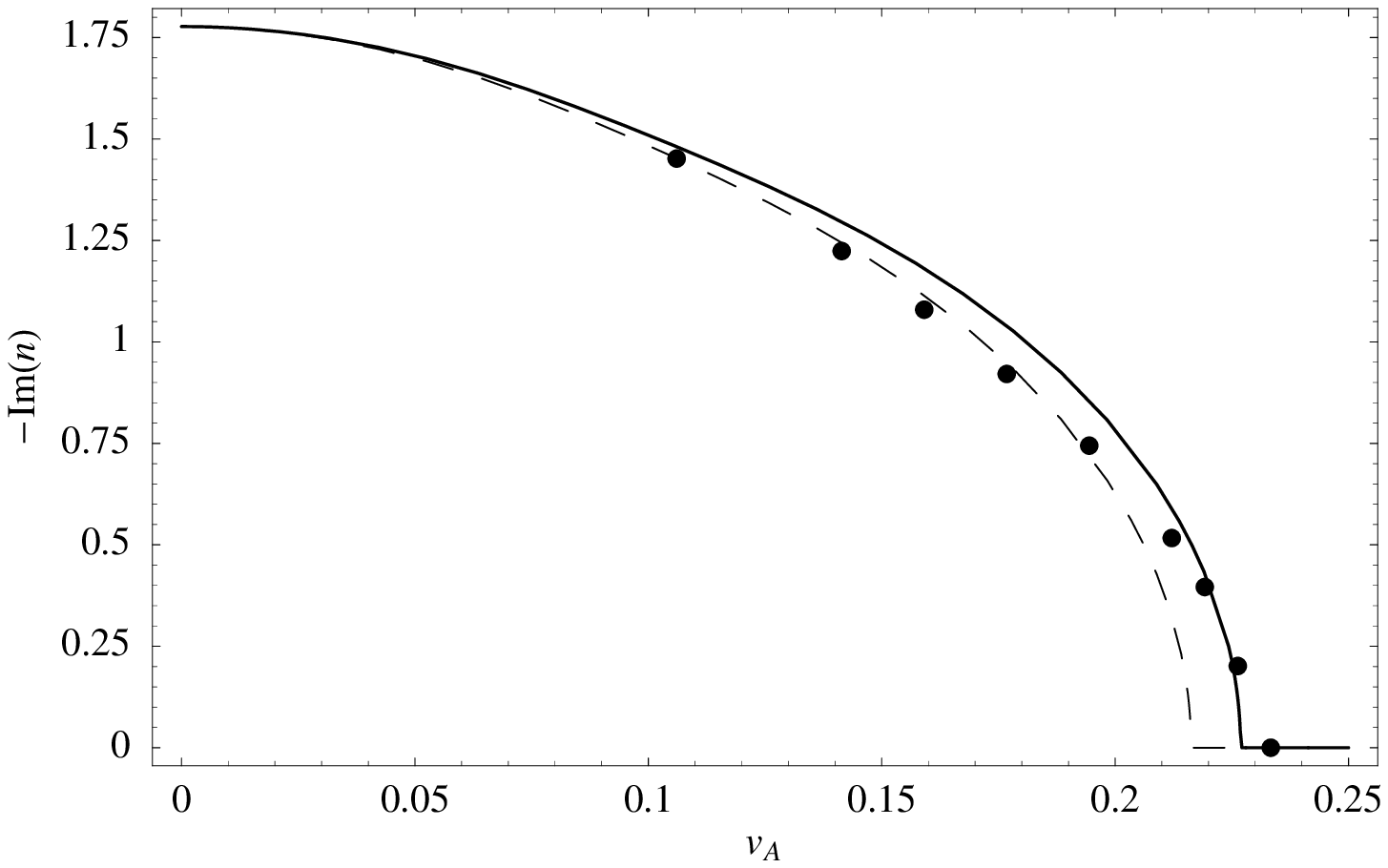}
\caption{Kelvin-Helmholtz growth rates as a function of magnetic field strength
(given here in terms of Alfv\'en speed, $v_A$) in the magnetized
layer, measured from simulation (diamonds) and predicted from
analytic theory (lines).  The dashed line gives the prediction using
the nominal width of the magnetized field layer, and the solid line
gives the prediction for the actual width of the magnetized layer
in the simulations after diffusion eats away at the profile as shown
in Fig.~\ref{fig:bprofile}.   Below $v_A \approx 0.1$, smaller-wavelength
modes, seeded by numerical grid noise, grow very quickly and contaminate
any attempt to measure a linear growth rate for the mode under consideration.
The theoretical curves are taken from numerical solution of the dispersion relation, Eq.~\ref{eq:khdispersion}.
}
\label{fig:khgrowth}
\end{figure}

The comparison of measured growth rates to analytic results is shown
in Figs.~\ref{fig:rtgrowth} and \ref{fig:khgrowth}, with theoretical results taken from the numerical
solution of Eqs.~\ref{eq:rtdispersion} and \ref{eq:khdispersion}.   Where the numerical
experiments can give clean growth rate measurements, the analytic and numerical
results agree to within a few percent, and there is also agreement at the
few percent level on the boundary of stability.

\section{DISCUSSION}
\label{sec:discussion}

We have shown, through derivation of dispersion relations and
two-dimensional numerical experiments, that it is possible for even thin
magnetized layers to  suppress instability growth on scales much larger
than their own thickness as long as the magnetic field strength is high
enough that the Alfv\'en velocity in the layer in the direction of the
perturbation is of order the relevant destabilizing velocity scales.
In the case of the Kelvin-Helmholtz instability, the most most relevant
case for core mergers, a magnetized layer can stabilize modes an order
of magnitude larger than the thickness of the layer if the Alfv\'en
speed of the same magnitude as the full shear velocity $v_A \approx 2 U$.
But this is almost automatically true; as shown in \cite{lyutikovdraping},
near the stagnation line the magnetic pressure reaches equipartition
with the ram pressure, meaning that this condition on the velocities
is met.   Thus we would expect, certainly near the stagnation line,
that the magnetic draping `protects' a merging core from instabilities,
as would appear to be the case in simulations published in the literature
\citep[][Dursi and Pfrommer, in preparation]{asai04, asai05, asai06}.

Clearly, this stabilization only applies to perturbations along the
field; but in this plane, then, the draped field can then have the twin
effects of protecting the merging core against thermal disruption, and
reducing the shear effects which would tend to mix in the core material
earlier.  In the plane perpendicular to the magnetic field, shear-driven
instability will occur unimpeded, leading to a distinct asymmetry in the
resulting magnetic layer and, presumably, the moving core or bubble.
Such asymmetries have been seen in 3d simulations in the literature,
as cited above, and would potentially be observable.  It is interesting
to note however that even a quite weak field can effect global mixing
in the presence of similar instabilities \citeeg{athenamhdrt}, and so
even in the plane perpendicular to the field it is possible that a thin
magnetized layer could keep an interface sharper than could exist absent
the magnetic field.

The full three dimensional stability problem remains to be tackled,
and less clear still is the effect of a more realistic magnetic
field, not expected to be planar or uniform, and the effects of fully
three-dimensional perturbations on such a layer.   Consideration of this
more complicated and realistic case is left to future work.

\bigskip 

\begin{acknowledgments}
LJD is grateful for discussions with M. Ruszkowski, K. Subramanian,
C. Pfrommer, and M. Lyutikov, which greatly contributed to this work,
for helpful comments on the manuscript by C. Pfrommer, A. Calder, and
M. Zingale,  and for the close reading and helpful suggestions and
corrections made by the anonymous referee.  The author acknowledges
funding from the National Science and Engineering Research Council,
the hospitality of W. Hillebrandt at the Max-Planck-Institut f\"ur
Astrophysik during the beginning of this work, and the hospitality
of the Kavli Institute for Theoretical Physics during its completion,
during which this research was supported in part by the National Science
Foundation under Grant No. PHY05-51164.  All computations were performed
on CITA's McKenzie and Sunnyvale clusters which are funded by the Canada
Foundation for Innovation, the Ontario Innovation Trust, and the Ontario
Research Fund for Research Infrastructure.  Simulations were performed
with version 3.0 of the Athena code.   Most of the grungy algebra was done
with Mathematica.  This work made use of NASA's Astrophysical Data System.
\end{acknowledgments}

\bibliographystyle{plainnat}
\bibliography{magneticdraping}

\clearpage

\end{document}